\documentclass[sn-nature]{sn-jnl}

\usepackage{graphicx}%
\usepackage[percent]{overpic}%
\usepackage{graphics}%
\usepackage{caption}%
\usepackage{adjustbox}%
\usepackage{epstopdf}%
\usepackage{multirow}%
\usepackage{amsmath,amssymb,amsfonts}%
\usepackage{amsthm}%
\usepackage{mathrsfs}%
\usepackage[title]{appendix}%
\usepackage{xcolor}%
\usepackage{textcomp}%
\usepackage{manyfoot}%
\usepackage{booktabs}%
\usepackage{algorithm}%
\usepackage{algorithmicx}%
\usepackage{algpseudocode}%
\usepackage{listings}%
\usepackage{graphicx}%
\usepackage{subcaption}%
\usepackage[utf8]{inputenc}
\DeclareUnicodeCharacter{2212}{-}
\DeclareUnicodeCharacter{03B4}{\ensuremath{\delta}}

\theoremstyle{thmstyleone}%
\theoremstyle{thmstyletwo}%

\theoremstyle{thmstylethree}%

\begin{document}
\title[Revealing the Influence of Dopants on the Properties of Fluorite Structure Ferroelectrics]{Revealing the Influence of Dopants on the Properties of Fluorite Structure Ferroelectrics}


\author*[1,2]{\fnm{Shouzhuo} \sur{Yang}}\email{shouzhuo.yang@ipms.fraunhofer.de}

\author[1]{\fnm{David} \sur{Lehninger}}

\author[1]{\fnm{Markus} \sur{Neuber}}

\author[1]{\fnm{Amir} \sur{Pourjafar}}

\author[1]{\fnm{Ayse} \sur{Sünbül}}

\author[1]{\fnm{Anant} \sur{Rastogi}}

\author[1]{\fnm{Peter} \sur{Reinig}}

\author[1]{\fnm{Konrad} \sur{Seidel}}

\author[1]{\fnm{Maximilian} \sur{Lederer}}

\affil*[1]{\orgdiv{Center Nanoelectronic Technologies}, \orgname{Fraunhofer IPMS}, \orgaddress{\street{An der Bartlake 5}, \city{Dresden}, \postcode{01109}, \state{Saxony}, \country{Germany}}}

\affil[2]{\orgdiv{Faculty of Electrical and Computer Engineering}, \orgname{Technische Universität Dresden}, \orgaddress{\street{Helmholtzstr. 10}, \city{Dresden}, \postcode{01069}, \state{Saxony}, \country{Germany}}}

\abstract{Fluorite structure ferroelectrics, especially hafnium oxide, are widely investigated for their application in non-volatile memories, sensors, actuators, RF devices and energy harvesters. Due to the metastable nature of the ferroelectric phase in these materials, dopants and process parameters need to be optimized for its stabilization. Here, we present clear evidence of how dopants affect the properties in this material system and solutions to achieve improved reliability, desired crystallization behavior and polarization hysteresis shape/position through co-doping. Finally, the benefits of co-doping in a variety of application fields are demonstrated.}


\maketitle

\section{Introduction}\label{sec1}
\begin{figure}[htb]
	\centering
	\includegraphics[width=\linewidth]{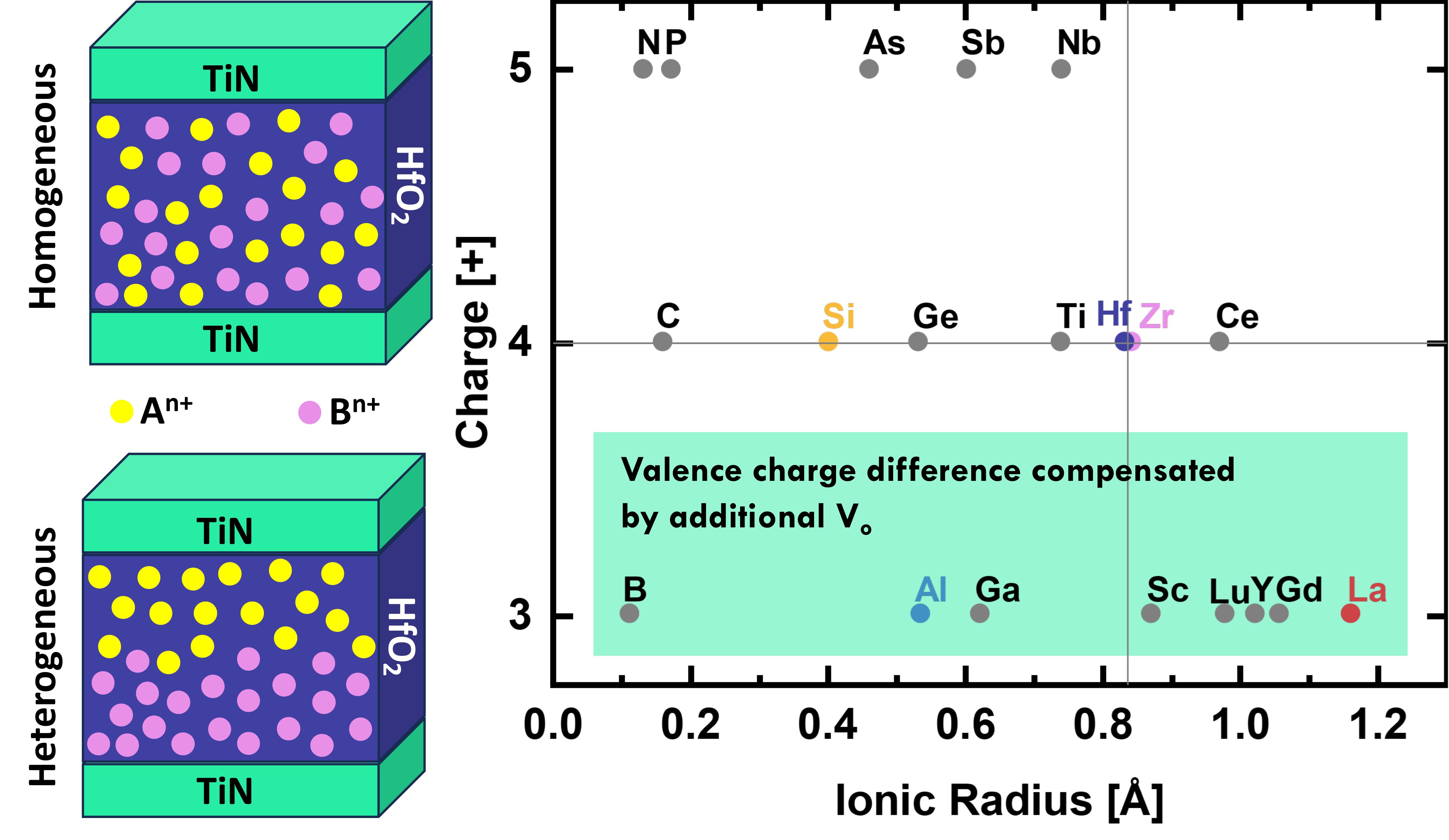}
	\caption{Schematic of homogeneous/ heterogeneous co-doping of HfO\textsubscript{2} and ionic radius of doping elements in hafnium oxide films in comparison to Hf. Dopants with lower or higher valence charge than Hf$^{4+}$ / Zr$^{4+}$ will alter the oxygen vacancy concentration in in hafnium oxide films. Values in the figure are taken from the work of R. D. Shannon \cite{ShannonACSA1976}.}
	\label{fig0}
\end{figure}

The ferroelectric phase of hafnium and zirconium oxide (HfO\textsubscript{2}/ZrO\textsubscript{2}) thin films is metastable in nature. Previous research has studied this behavior extensively and provided insights into the thermodynamics and kinetics of this material system. \cite{Chae2020JAP,Fields2022AEM,Park2019145FDHO,Park2019AEM} Density functional theory based calculations have revealed that mechanical stress and doping strongly influence the thermodynamic energy levels of the crystallographic phases of hafnium and zirconium oxides. \cite{Künneth2019245FDHO,Chen2019MRE,Mukherjee2024NPJ} The ferroelectric phase has been experimentally confirmed to be the orthorhombic phase ($Pca2_1$) \cite{Mukundan2020APL,Shimizu2021PSSRRL}, rendering HfO\textsubscript{2}- and ZrO\textsubscript{2}-based materials as ferroelectrics with a fluorite structure. As Park et al. illustrated \cite{Park2019Nanoscale}, this phase is stabilized through kinetical means. Here, the nucleation begins in the tetragonal phase. The transition into the monoclinic phase upon cool down is kinetically suppressed, e.g. by thermal quenching \cite{Wen2024JAP,Narasimhan2021PSSRRL} or a capping layer \cite{Fields2022AEM,Lee2019AEM}, and the $Pca2_1$ phase emerges below the Curie temperature. Recent studies on the microstructure of HfO\textsubscript{2} films are in agreement with this process but have found evidence for possible nucleation of monoclinic phase as well \cite{Yang2024APL}. The crystallographic properties of the substrate or seed layer are therefore of major importance for the resulting phase composition as well as microstructure \cite{Yang2024APL,Lederer2021ACSAEM}.\\
The excellent ferroelectric properties of this material system as well as its CMOS-compatibility lead to a strong interest for its integration into nanoelectronic applications, ranging from memories and artificial intelligence to sensors and RF devices. Nevertheless, reliability challenges currently remain, hindering the large-scale commercialization of these ferroelectric devices. Endurance and imprint are particularly major concerns in this material system. \cite{Park2018795MRS,Mulaosmanovic2021IRPS} Previous research has already highlighted the importance of defects, e.g. oxygen vacancies, as well as microstructure on reliability aspects \cite{Bao20234615ACSAEM,Yuan20223667NanoR}, but an in-depth understanding is still missing.\\ This work demonstrates how the crystallization behavior, the ferroelectric hysteresis response and the reliability performance can be precisely controlled independently through the means of co-doping. This approach uses the insertion of multiple impurity elements, to control oxygen vacancy levels and mechanical microstress (see Fig. \ref{fig0}). Based on the valence charge and ionic radius of possible elements these properties can be altered without leaving the ideal doping concentration for the ferroelectric phase.\\ 
In addition, the scalability of this approach is demonstrated through the integration into established industrial technology nodes. The resulting ferroelectric performance and reliability of the integrated devices for Ferroelectric-Metal-Field-Effect Transistor (FeMFET) and Ferroelectric Random Access Memory (FeRAM) applications is significantly enhanced. Moreover, advantages for pyroelectric sensors through co-doping are demonstrated as well.\\

\section{Crystallization behavior engineering}\label{sec2}
The crystallization of HfO\textsubscript{2}/ZrO\textsubscript{2} thin films is of major importance for the application, as the thermal budget of front-end-of-line (FEoL) and back-end-of-line (BEoL) CMOS-nodes may result in film degradation \cite{Nagasato200831JJAP,Kong2019Coatings} or insufficient crystallization \cite{Toyoda2009JAP,Cho2024IEEEJEDS,Hsain2020APL}, respectively. In addition, the resulting microstructure will affect the variability in scaled devices and the switching behavior through the orientation of the polarization axis \cite{Lee2022ActaMat, Margolin2025ActaMat,Lederer2022FN,Lederer2022PSSRRL}. The next sections will therefore deal with the impact of homogeneous co-doping on the crystallization temperature (T\textsubscript{cryst}) as well as local effects of heterogeneous co-doping based on the dopant position for the crystallization and nucleation processes.\\

\subsection{Control over crystallization temperature}
Previous research has evidenced that the T\textsubscript{cryst} of ferroelectric hafnia and zirconia thin films is highly dependent on many factors, the film thickness \cite{Lehninger2020PSSa,Shiraishi2017239MSSP}, the dopants concentration \cite{Lehninger2021AEM,Tashiro20213123ACSAEM}, and the integrated stack (MFM/MFIS) \cite{Lederer20214115ACSAEM,Qi2021APL}. For instance, a 10~nm HfO\textsubscript{2} thin film doped with equal-ratio zirconium (HZO) in an MFM stack can be induced in ferroelectric phase below 400~°C \cite{Lehninger2020ISAF}. In contrast, the HfO\textsubscript{2} doped with aluminum (Al) or silicon (Si) requires a much higher temperature for crystallization.\\
It is demonstrated in Fig. \ref{fig1}a, that the T\textsubscript{cryst} of HZO thin film can be almost linearly varied via homogeneous co-doping of Al/Si into HZO thin films (HZAO/HZSO). The generality of this trend is further verified through HZSO co-doped thin films with varying thicknesses, as depicted in Fig. \ref{fig1}b. Additionally, at each HZSO ratio, T\textsubscript{cryst} increases with a decrease in film thickness. The raw data of grazing incidence X-ray diffraction (GIXRD) for determination of T\textsubscript{cryst} of each sample can be found in Fig. S1 and S2 of supplementary materials.\\
Consequently, the compatibility of co-doped HfO\textsubscript{2} thin films with BEoL can be anticipated and controlled.\\

\subsection{Control over nucleation process}
\begin{figure}[htb]
	\centering
	\includegraphics[width=\linewidth]{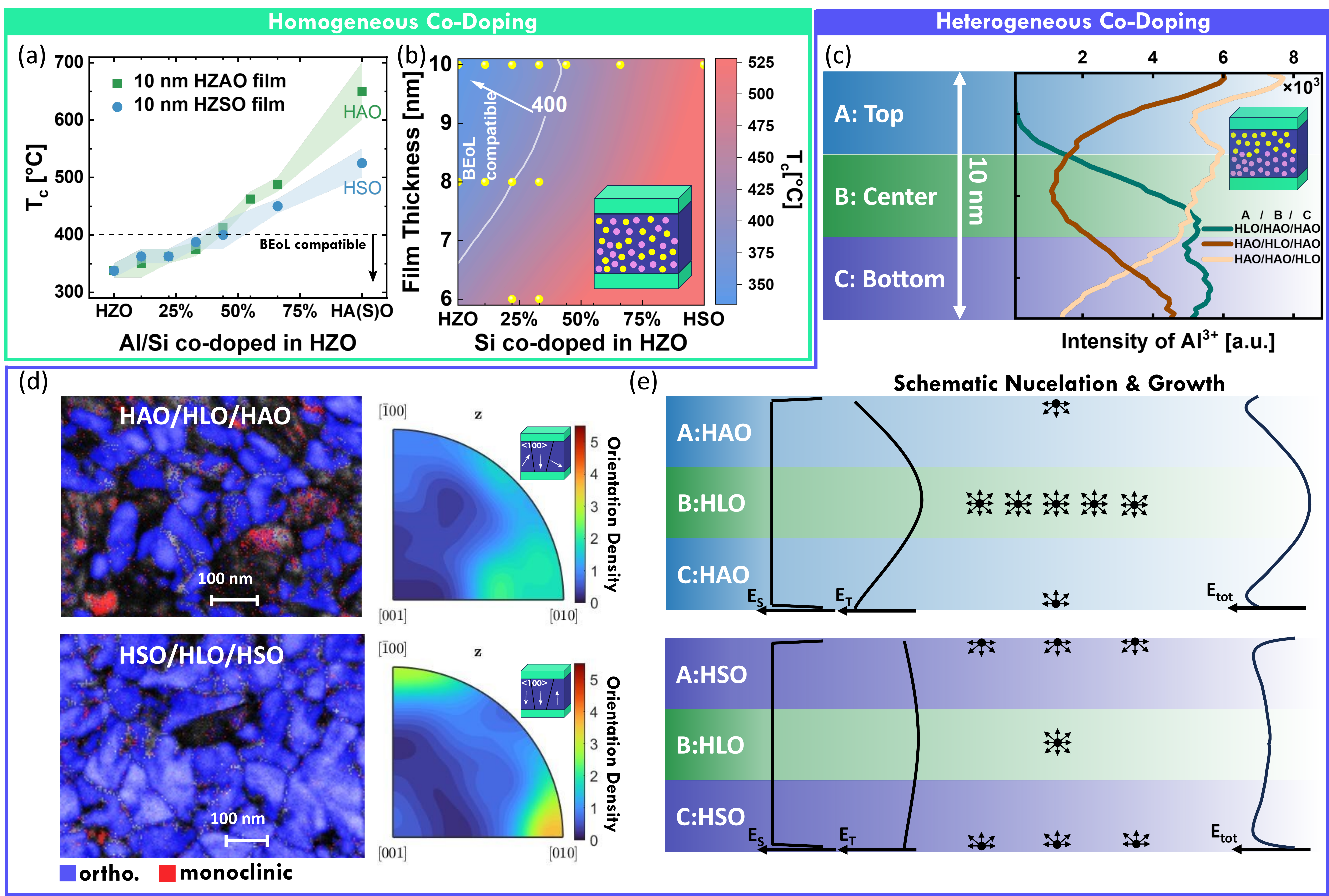}
	\caption{(a,b) Crystallization temperature (T\textsubscript{cryst}) of homogeneously co-doped HfO\textsubscript{2} thin films: (a) 10~nm HZAO and HZSO, (b) HZSO with varying dopant ratios and thicknesses, (c) ToF-SIMS results of three samples heterogeneously co-doped with Al and La, (d) TKD analysis and the corresponding IPFs of HAO/HLO/HAO as well as HSO/HLO/HSO stacks, (e) mechanism of phase and texture control of poly-crystalline HfO\textsubscript{2} film.}
	\label{fig1}
\end{figure}

The preceding section highlights that the homogeneous co-doping method can effectively control the T\textsubscript{cryst} of HfO\textsubscript{2} thin films in MFM stack by incorporating and blending various dopants while adjusting their ratios.\\
Beyond homogeneous co-doping, another co-doping method, termed heterogeneous co-doping, is schematically depicted in Fig. \ref{fig1}c. Unlike homogeneous co-doping, where both dopants are uniformly distributed throughout the thin film, heterogeneous co-doping involves two dopants distributed separately in n-distinct layers, as shown here for n=3 (A, B, and C) within e.g. a 10~nm HfO\textsubscript{2} thin films. A series of MFM stacks with 10~nm lanthanum (La) and Al heterogeneous co-doped hafnia (HLAO) are fabricated, and the results from time-of-flight secondary ion mass spectrometry (ToF-SIMS) indicate expected vertical distributions of La and Al atoms.\\
Following the annealing process at 800°C for 20 s, the results from GIXRD, as depicted in Fig. S3, reveal a significant dependence of crystalline phase and orientation on the distribution of dopants. This dependence contributes to the much lower T\textsubscript{cryst} of HLO \cite{Kozodaev2017APL} compared to HAO, thereby affecting the energy barrier of nucleation within the HfO\textsubscript{2} thin films. Specifically, when La is doped in layer B, more nuclei are driven to form at the center of the film. Consequently, a much more intensive monoclinic phase is detected due to the absence of mechanical confinement from electrodes \cite{CHERNIKOVA2015ME} during grain growth.\\

Transmission Kikuchi diffraction (TKD) also confirmed the presence of the monoclinic phase marked in red, as illustrated in Fig. \ref{fig1}d. It is noted that the grains in blue are orthorhombic and in black are not determined. Furthermore, the corresponding inverse pole figure (IPF) also exhibits no distinct texture due to the random growth of grains from the center of the film. In contrast, when Al is replaced by Si, the T\textsubscript{cryst} of HLO is not low enough to force the majority of nuclei formed at the center. Consequently, most of the grains grow from the interfaces with the electrodes, where the lowest energy barrier is observed, leading to confinement and crystallographic texture originating from the electrodes (as illustrated in Fig. \ref{fig1}e).\\

\section{Ferroelectric hysteresis engineering}\label{sec3}
Besides influencing crystallization behavior, co-doping allows precise control over the hysteresis shape. Mechanical stress \cite{Shiraishi2016APL} and crystallization behavior affect hysteresis, as does the distribution of oxygen vacancies. It is therefore essential to control these different contributions separately.\\
\subsection{Impact of crystallographic texture}
\begin{figure}[htb]
	\centering
	\includegraphics[width=\linewidth]{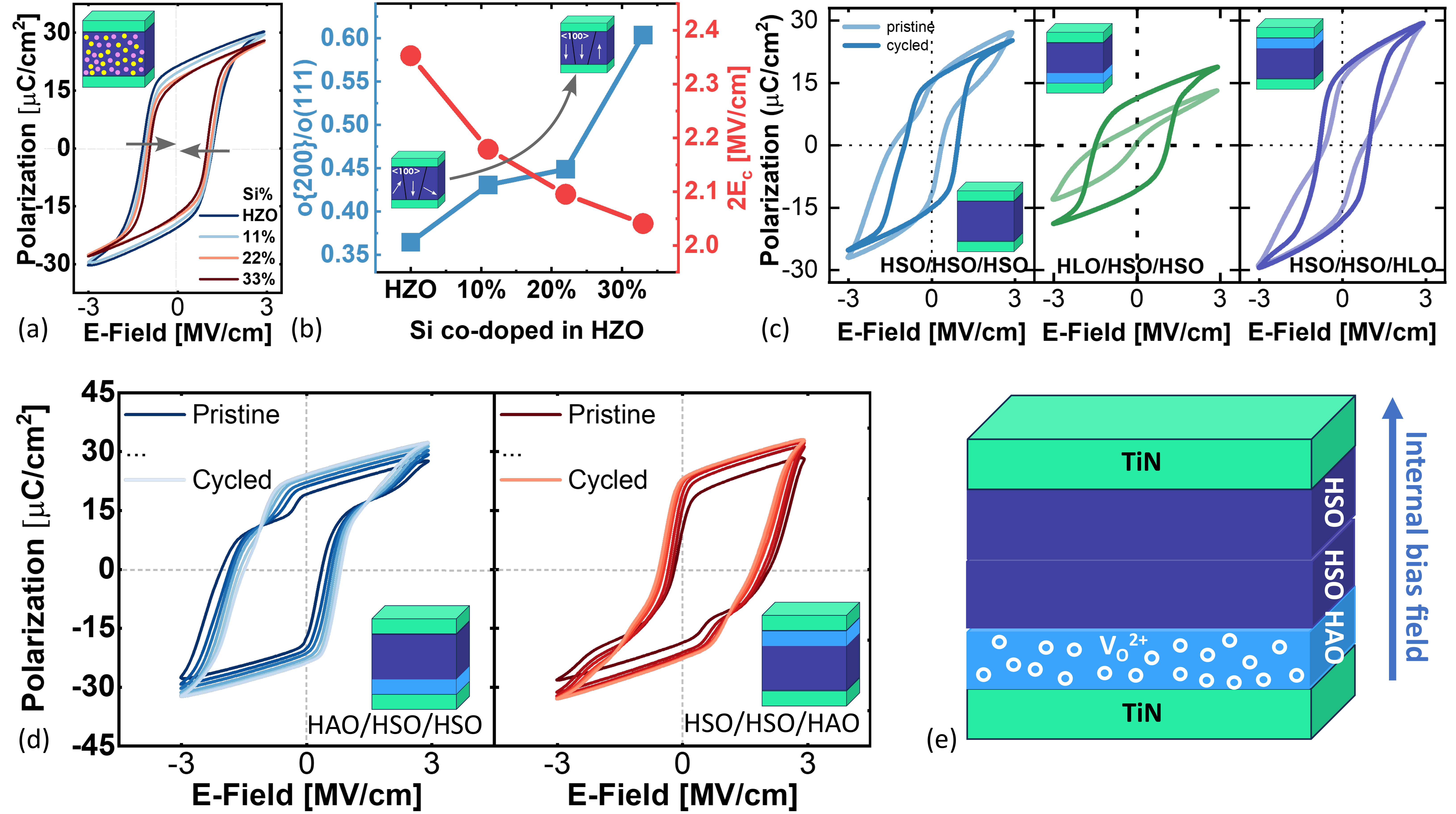}
	\caption{(a) P-E loops of homogeneously co-doped HZSO thin films with various Si ratios, (b) intensity ratio of o\{200\} and o(111) and the corresponding E\textsubscript{C} of homogeneously co-doped HZSO thin films with various Si ratios, (c) P-E loops of HLO heterogeneously co-doped in HSO thin films with various HLO location, (d) P-E loops of HAO heterogeneously co-doped at bottom and on top in HSO thin films, and (e) schematic of induced internal bias field within HfO\textsubscript{2} by oxygen vacancies (V${_o}$$^{2+}$) that introduced by HAO in a HSO/HAO heterogeneously co-doped thin film.}
	\label{fig2}
\end{figure}

As shown in previous works, the crystallographic orientation of the material can be tuned through the substrate and semi-epitaxial growth \cite{Lederer2021pssRRL,Lederer2021APL}. As shown for HSO, the semi-epitaxial growth window on TiN requires annealing temperatures in the range of 800~°C to 950~°C \cite{Lederer2021APL}. In HZO, the required temperatures already lead to a degradation, e.g. increased leakage and significantly reduced endurance. \cite{Asapu2022FrontMat}\\
Utilizing Al or Si homogeneous co-doping, semi-epitaxial growth can be achieved in HZO without degradation, as shown in Fig. S4 of supplementary materials. As a result, very square-like hysteresis with low coercive field (E\textsubscript{C}) can be observed in Fig. \ref{fig2}a. This is a direct consequence of the crystallization processes discussed in section \ref{sec2}, since Al and Si homogeneous co-doping will increase the T\textsubscript{cryst}. While heterogeneous co-doping would allow to control this more precisely, already homogeneous co-doping can achieve this within the annealing temperature window of 650~°C to 800~°C for 2.4 - 3 at.\% Al. \cite{SUNBUL2024Elsevier}\\
By controlling the crystallographic texture of the film, the electrical behavior can be modified. As shown in Fig. \ref{fig2}b, the E\textsubscript{C} decreases with increasing Al-/Si-content. In addition, the line intensity of the o\{200\} lines in the GIXRD measurement increases relative to the intensity of the o\{111\} line, confirming the increasing contribution of semi-epitaxial growth during the crystallization of the film.\\ 
As shown in Fig. \ref{fig2}a, the decrease in E\textsubscript{C} is caused solely by an increasing slope of the hysteresis flanks and not an increase in antiferroelectric-like behavior. This originates from a narrow E\textsubscript{C} distribution, which is a direct result of the narrow orientation distribution of polarization axis due to the semi-epitaxial growth. The hysteresis response can therefore be adapted without changing the annealing temperature, allowing to tune between analogue- and digital-like switching (broad and narrow E\textsubscript{C} distribution, respectively) as needed by the application.\\
\subsection{Oxygen vacancy engineering}\label{sec:Vo}
As discussed in earlier sections, dopants differ in valence charge (see Fig. \ref{fig0}). As a consequence, distribution of oxygen vacancies can be influenced by selecting certain dopants.
For example, as illustrated in Fig. \ref{fig2}c, the P-E loops of the Si mono-doped HfO\textsubscript{2} exhibit an imprint at pristine state in the negative direction, attributed to distinct processes occurring at the top (PVD) and bottom (ALD) TiN electrodes, resulting in an increased oxygen vacancy concentration at the TiN-HfO\textsubscript{2} interfaces. In the co-doped samples, the imprint is intensified when La is doped at the bottom of the HfO\textsubscript{2} film. Conversely, when La is doped at the top of HfO\textsubscript{2}, the imprint is effectively "corrected".\\
To clarify the impact of oxygen vacancies, heterogeneous co-doping of Si and Al in HfO\textsubscript{2} is explored. Like shown in Fig. \ref{fig0}, Si and Al are quite similar in their ionic radius. \cite{ShannonACSA1976} Consequently, no significant differences in the induced microstress are expected. On the other hand, Al is, like La, of the charge 3+. In consequence, introduction of oxygen vacancies with Al-doping is expected. Since both dopants have a similar doping optimum, it further enables the substitution of Si-doping cycles with Al, and vice versa. Therefore, the location of oxygen vacancies in the HfO\textsubscript{2} layer can be controlled precisely.\\
Comparing two stacks with symmetrical stacks, namely HSO sandwiched in between HAO and vice versa, no significant difference between the two layers is observable (see Fig. S5 in supplementary materials).\\
By investigating the two extreme cases of HAO deposited only at the top or bottom of the layer, respectively, the influence of the local arrangement of the oxygen vacancies can be probed. From the polarization hysteresis in Fig. \ref{fig2}d, it is clear that the location of the HAO at the top or bottom induces a location dependent imprint in the capacitor. Since the pulses are applied from the bottom of the stack, the polarity of the imprint agrees well with the internal bias field expected for the localized oxygen vacancies (see Fig. \ref{fig2}e). 
 
\section{Reliability engineering HZAO}\label{sec4}
\begin{figure}[htb]
	\centering
	\includegraphics[width=\linewidth]{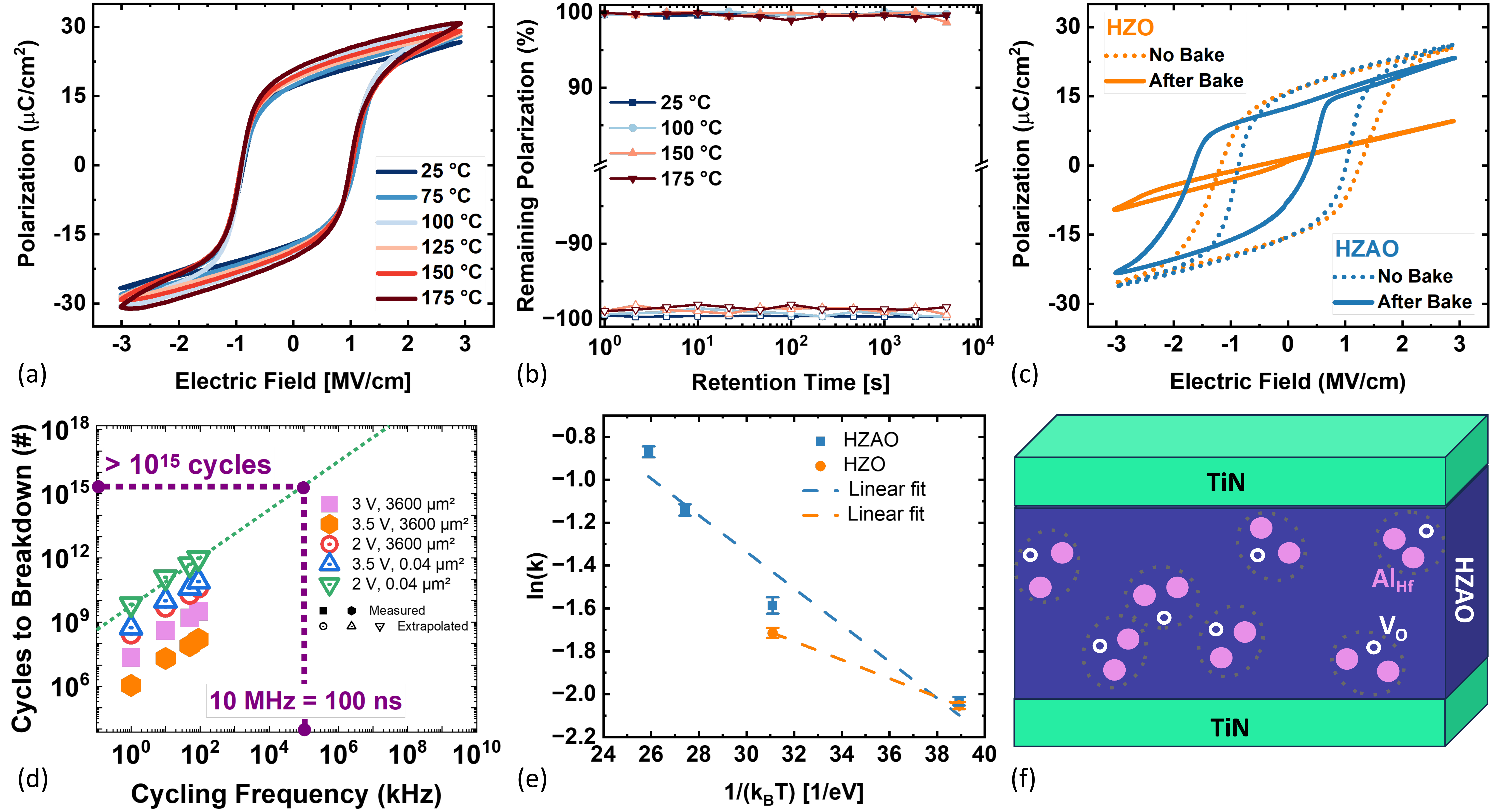}
	\caption{(a) P-E loops of homogeneously co-doped HZAO until 175~°C, (b) retention of homogeneously co-doped HZAO until 175~°C, (c) Imprint characteristics of HZO and HZAO: Initial P-E loops (dashed line) and after 12-hour temperature stress at 150~°C (solid line), (d) Arrhenius plots of mono-doped HZO and homogeneously co-doped HZAO, and (e) endurance of homogeneously co-doped HZAO extended to 100 kHz for cycling \cite{Sünbül2024IRPS}, and (f) Schematic of local binding oxygen vacancies in homogeneously co-doped HZAO film}
	\label{fig3}
\end{figure}

Oxygen vacancies have furthermore been identified as a crucial parameter for the device reliability, especially endurance, imprint and retention \cite{Ali2024APL}. Previous works have suggested that they play an essential role during the electric field cycling of the device, influencing the wake-up effect \cite{Chouprik2019ACSAEM} as well as the fatigue \cite{Bao2023ACSAEM}. In both cases, the redistribution of these defects is the major contributing factor, but some works have also suggested oxygen vacancy generation at the interfaces \cite{Goh2020IEEETED,Barrett2024APL}.\\
In case of the wake-up effect which describes the transition from a pinched or antiferroelectric-like hysteresis toward a square-shaped ferroelectric hysteresis, oxygen vacancy redistribution can lead to domain depinning \cite{Starschich2016APL,Fengler2019FeBook,Zhang2024IEEETED} as well as a reduction of imprint \cite{Fengler2018JAP,Chen2023NanoScale}. Here, the oxygen vacancies are moving from the interfaces into the bulk leading to a more homogeneous distribution. Especially the case of imprint and how to circumvent it has been discussed in detail in section \ref{sec:Vo}. For fatigue, on the other hand, oxygen vacancies redistribute to form filaments, analogue to resistive random access memory based on hafnium oxide. In consequence, these filaments act as leakage pathways, increasing leakage current, and result in the worst case in a device breakdown.\\
In addition, retention of ferroelectric capacitors is suffering from imprint losses due to pinning of domains through oxygen vacancy redistribution. Here, oxygen vacancies move to the domain interfaces and act as a local pinning side. \cite{Fengler2019FeBook,Fan2024TSF}\\
To clarify the influence of the oxygen vacancies introduced by trivalent dopants on the reliability, it is therefore investigated by retention and endurance measurements. While a reduction of endurance and retention would be expected with a higher amount of oxygen vacancies (due to the oxygen vacancy redistribution, as discussed above), no such trend is observed when increasing the amount of co-doping with trivalent dopants. On contrary, an improvement in all reliability aspects can be observed: i) No significant change of P-E curves with temperature is observable (Fig. \ref{fig3}a). ii) No significant loss in retention is observable over the measured timescale (Fig. \ref{fig3}b). iii) The rate of imprint is strongly reduced compared to regular HZO (Fig. \ref{fig3}c). iv) Endurance is significantly enhanced, suggesting $>10^{15}$ cycles \cite{Sünbül2024IRPS} (Fig. \ref{fig3}d). 
\\
Consequently, co-doping with trivalent dopants leads to drastic improvement in reliability, making fluorite structure ferroelectrics compatible with automotive and industrial requirements, such as AEC-Q100 \cite{11026934}. The cause for this can be explained by the reduced rate of imprint over time. When comparing the rate at different temperatures, Arrhenius behavior can be observed for both cases, HZO and HZAO, in Fig. \ref{fig3}e. More details on this figure of merit is presented in Fig.  S6 of supplementary materials. The extracted effective activation energy barriers differ significantly, with approximately 0.043~eV and 0.085~eV for HZO and HZAO, respectively. This is likely caused by a local binding of the oxygen vacancies in proximity to the trivalent dopant sites, as presented in Fig. \ref{fig3}f. As a result, oxygen vacancy movement is strongly suppressed, preventing redistribution in case of fatigue and retention.\\

\section{Enabling new and improved devices}\label{sec5}
\subsection{Ferroelectric BEoL Memory Concepts}
The MFM stack, discussed thus far, serves as a fundamental component for FE memories. Situated within the BEoL, it can be connected either to the gate or the drain contact of a standard logic device, facilitating the creation of a FeMFET or a FeRAM, respectively (illustrated in Fig. \ref{fig5}a).\\
In the FeMFET, the polarization state of the BEoL MFM capacitor modulates the surface potential of the channel, influencing the threshold voltage (V$_{th}$) of the transistor. Readout occurs by sensing the drain current at low gate voltages (well below the coercive field E\textsubscript{C}). Readout is thus non-destructive.\\
Despite successful demonstrations of FeFETs (i.e., 1T memory cells without the BEoL MFM-module) in 28~nm and 22~nm CMOS technology, achieving transistor-level integration poses challenges such as the depolarization field at the FE/insulator interface in the MFIS configuration and the integration of FeFET memory cells with logic devices. The FeMFET concept aims to address these challenges by replacing the MFIS stack with an MFM stack. Furthermore, isolating the MFM module from the logic level opens new avenues for optimizing a memory cell by independently tuning the transistor and MFM module.\\
In contrast to the FeMFET, the readout process for the 1T1C FeRAM is destructive, alike to a DRAM. However, unlike DRAM, which stores information as charges (volatile), FeRAM utilizes the remanent polarization state of the FE material, rendering it non-volatile. During readout, the FE polarization is manipulated, and the resulting displacement current is detected.\\
The electron micrograph in Fig. \ref{fig5}b portrays a square-shaped MFM module integrated between metallization planes M4 and M5 in the BEoL of XFAB's XT018 technology. This state-of-the-art 180~nm BCD-on-SOI technology is tailored for Grade 0 applications in the automotive industry according to the AEC-Q100 standard. Notably, all interfaces exhibit smooth surfaces, devoid of any signs of interface reactions.\\
Fig. \ref{fig5}c depicts the P-E hysteresis loop obtained from the BEoL-integrated MFM module utilizing the reliability-engineered HZAO material discussed in section~\ref{sec4} after undergoing 10$^4$ wake-up cycles. These hysteresis loops exhibit the anticipated values for 2P\textsubscript{r} and E\textsubscript{C}, which are essential for ensuring the optimal functionality of memory cells in the mentioned memory technologies.\\
\subsection{Pyroelectric devices for sensor applications}
The pyroelectric properties of doped HfO\textsubscript{2} can be used for various applications, like energy harvesting, cooling and for sensor applications. Especially pyroelectric sensors have already been introduced and will gain even more market share in the next years. They can be used for temperature measurement, motion detection, gas analysis and flame
detection.\\
The common material for these applications is Lithium Tantalate with a pyroelectric coefficient of 170~$\mu$C/m$^2$/K, but it is not compatible with CMOS fabrication processes. HfO\textsubscript{2} on the other hand exhibits a significantly lower pyroelectric coefficient \cite{Jachalke2018APL}, but by depositing pyroelectric HfO\textsubscript{2} into 3D structures, the pyroelectric output can be multiplied by the aspect ratio of those structures \cite{Hanrahan2019EneTech}.\\
One other key for the optimization is the crystal phase of the HfO\textsubscript{2} films. It was found that the pyroelectric efficient in HfO\textsubscript{2} is strongest in the transition from an orthorhombic to a monoclinic/tetragonal state \cite{Mart2021APLMat}. By doping HfO\textsubscript{2} with multiple doping elements film stress and oxygen vacancies can be tuned, as discussed in section~\ref{sec4} and \ref{sec5}.\\
\begin{figure}[tb]
	\centering
	\includegraphics[width=\linewidth]{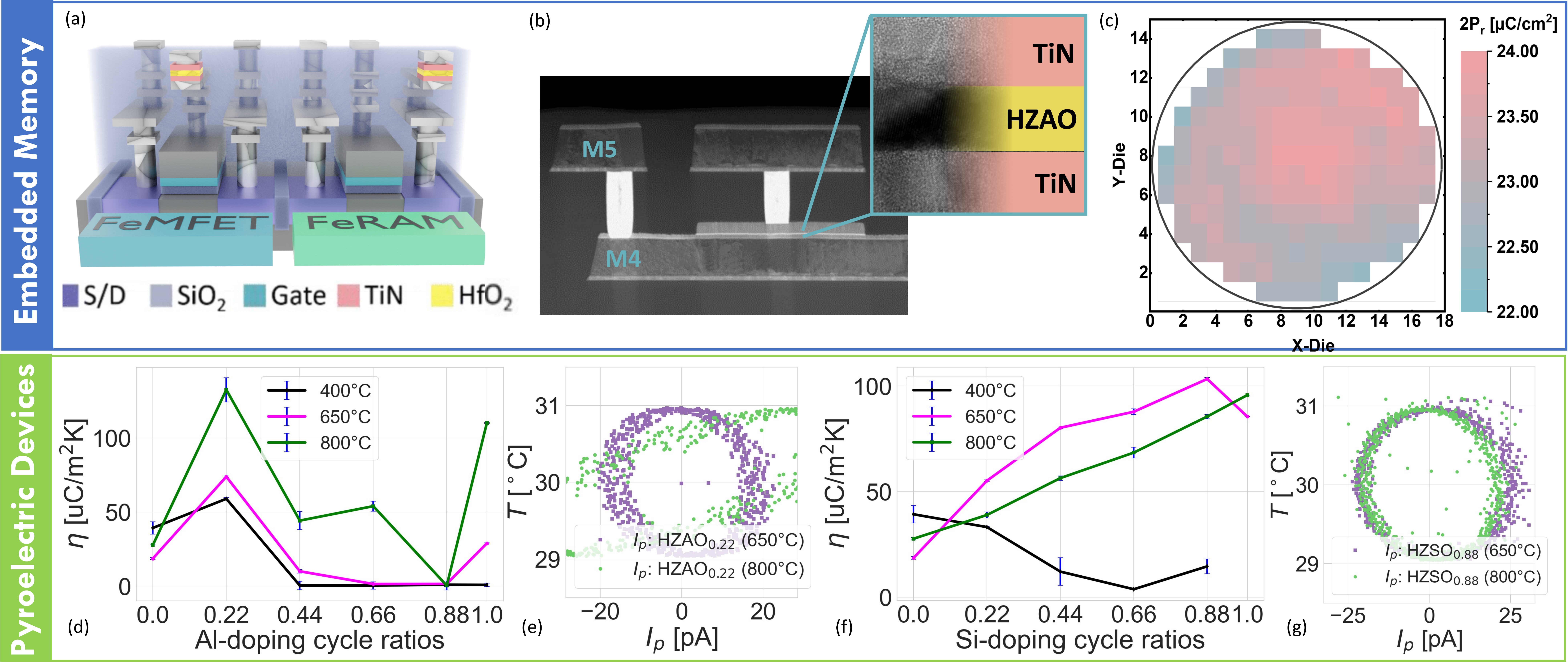}
	\caption{BEoL-integrated MFM Module and its characteristics: a) schematic of FeMFET and 1T1C FeRAM with BEoL MFM module, b) electron micrograph of the MFM module in the BEoL of CMOS technology, c) wafer-scale characterization of 2P\textsubscript{r} after wake-up, showing expected and homogeneous polarization over the full wafer, (d-g) pyroelectric coefficient and Lissajous plot of co-doped HZAO and HSAO from Sharp-Garn measurement.}
	\label{fig5}
\end{figure}
The samples in Fig. \ref{fig5} were co-doped by sequencing ZrO\textsubscript{2}, Al$_2$O$_3$ and SiO$_2$ ALD to form 10~nm thin films of HZAO and HZSO. The substitution of Zr with Al (Fig. \ref{fig5}d) and Si (Fig. \ref{fig5}f) influences the pyroelectric behavior and therefore also the necessary thermal treatment of the samples. Co-doping with Al results in a maximum of the 129.8~$\mu$C/m$^2$/K for samples annealed at 800~°C. This peak is also shown at the same Al ratio for samples which were treated with 400 and 650~°C. The Lissajous plot (Fig. \ref{fig5}e) of the represented sample suggest high influence of non-pyroelectric current like thermally induced charge detrapping \cite{Sharp1982JAP}. For HZSO the highest pyroelectric coefficient was measured for sample with an Si ALD cycle ratio of 0.88 and a thermal anneal at 650 °C. The data of the Lissajous measurement for this sample reveals a lower influence of noise on the pyroelectric measurement than witnessed in the HZAO sample.\\
These results promise a great potential for the optimization of the pyroelectric response in sensor devices. Furthermore, these results indicate that the presence of oxygen vacancies, introduced by the Al co-doping, play a vital role for achieving high pyroelectric coefficients in these fluorite ferroelectrics.\\
\section{Conclusion}\label{sec6}
In summary, the results presented here demonstrate the effectiveness of co-doping in ferroelectric hafnium oxide thin films to control phase composition, crystallization temperature, crystallographic texture and reliability. Using this approach, high temperature reliability, excellent pyroelectric and ferroelectric performance was demonstrated. Furthermore, it is compatible with industrial CMOS processes and the integration into XFAB's XT018 technology was successfully demonstrated.\\ 

\bibliography{sn-bibliography}

\section*{Acknowledgements}\label{sec7}
The authors would like to thank their colleagues at Fraunhofer IPMS-CNT for their valuable contributions. Fred Schöne support the deposition of co-doped HfO\textsubscript{2} films. Lisa Roy and Kati Biedermann performed the GIXRD measurements. Jennifer Salah Emara and Kati Biedermann carried out the ToF-SIMS measurements. André Reck conducted the TKD measurements. Raik Hoffmann supported electrical characterization of devices.\\
This work has received funding from the ECSEL Joint Undertaking (JU) project StorAIge under grant agreement No 101007321. The JU receives support from the European Union’s Horizon 2020 research and innovation programme and France, Belgium, Czech Republic, Germany, Italy, Sweden, Switzerland, Turkey, and funding from German Bundesministerium für Bildung und Forschung (BMBF) through the project T4T under grant agreement No. 16ME0483 and the project Polar under grant agreement No. 13N15141-13N15144.\\
\section*{Author information}\label{sec8}
\subsection*{Authors and Affiliations}
A. Pourjafar is now affiliated to Fraunhofer Institute for Ceramic Technologies and Systems IKTS.
\subsection*{Contributions}
S. Yang summarized the data, organized the manuscript, and wrote Section 2 under the supervision of M. Lederer. D. Lehninger, P. Reinig, M. Neuber and K. Seidel collected data, wrote Section 5, and reviewed the other sections. S. Yang and A. Pourjafar collected data for Section 3 and reviewed the other sections. A. Sünbül and D. Lehninger collected data and wrote Section 4. S. Yang and A. Rastogi collected data for Section 2 and reviewed the other sections. K. Seidel acquired and managed the projects and reviewed the entire manuscript. M. Lederer conceived the idea, proposed the article structure, wrote Sections 1, 3, and 6, and reviewed the entire manuscript.
\end{document}